\documentclass[9pt,conference]{IEEEtran}
\usepackage{amsmath,graphicx,url,times}
\usepackage{amssymb}
\usepackage{color}
 \usepackage{waspaa25} % removes hyperlinks (required by IEEE Xplore)

\usepackage{bm} % for bold math symbols (incl. Greek letters) with \bm{}

\title{Diverse Audio Embeddings-- \\ Bringing Features Back Outperforms CLAP !}

\name{Prateek Verma}
\address{Stanford University, Stanford, CA, USA \;
}
\usepackage{cuted}
\usepackage{capt-of}
\usepackage[symbol]{footmisc}

\begin{document}
\maketitle

\begin{sloppy}
\begin{strip}\centering
\vspace{-1.3cm}
\includegraphics[width=\linewidth,height=8cm,keepaspectratio]{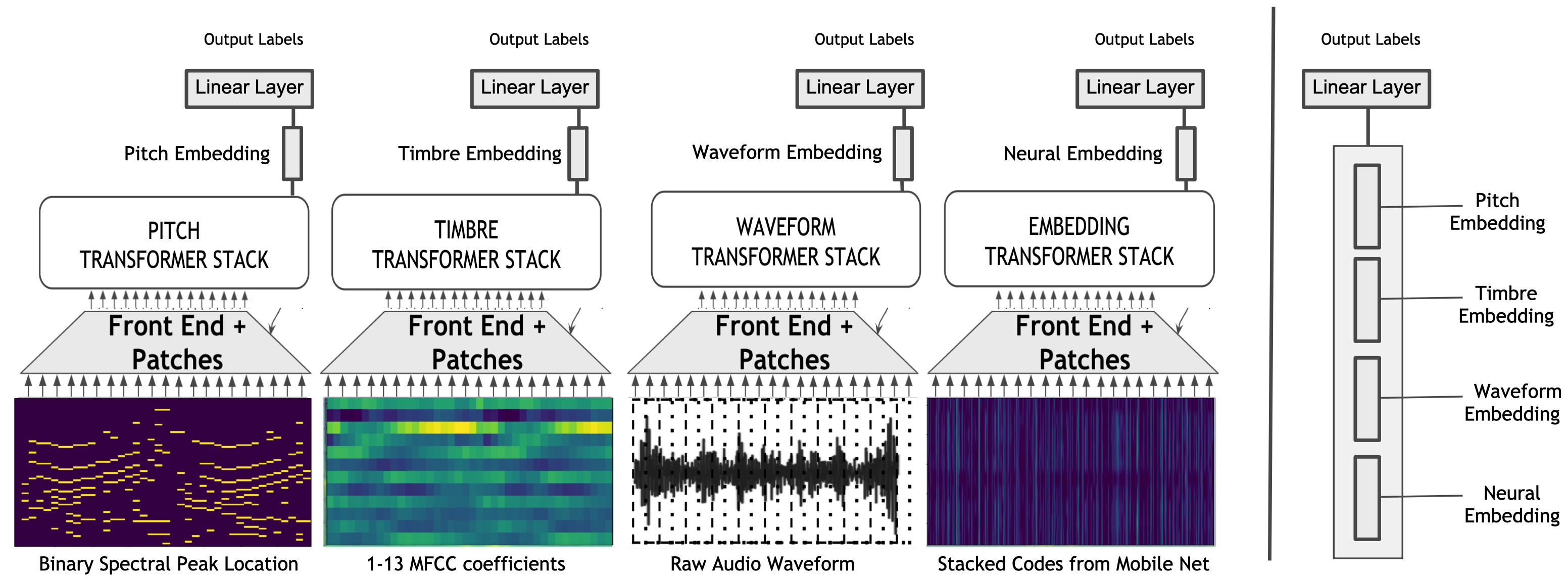}
\captionof{figure}{An overview of the proposed method
\label{fig:feature-graphic}}
\end{strip}
\end{sloppy}
\begin{abstract}
With the advent of modern AI architectures, a shift has happened towards end-to-end architectures. This pivot has led to neural architectures being trained without domain-specific biases/knowledge and optimized according to the task. In this paper, we learn audio embeddings via diverse feature representations, which, in this case, are domain-specific. For audio classification over hundreds of categories of sound, we learn robust separate embeddings for diverse audio properties such as pitch, timbre, and neural representation, including via an end-to-end architecture. We observe handcrafted embeddings, e.g., pitch and timbre-based, although they cannot beat a fully end-to-end representation. However, combining all of them together helps us significantly improve performance. This work would pave the way to bring some domain expertise with end-to-end models to learn robust, diverse representations, surpassing the performance of just training end-to-end models.
\end{abstract}\section{Introduction and Related Work}
We interact with sounds every day. They occur in various environments and places around us, with their diversity and richness described in \, \cite{gemmeke2017audio}, having the most extensive ontology of everyday sounds. Making computers hear similar to humans has come realistically close to achieving super-human performance, with the advent of transformer architectures \cite{vaswani2017attention}. They have not only revolutionized natural language processing \cite{vaswani2017attention,wei2021finetuned}, they also have altered the course of research in problems in areas such as computer vision \cite{dosovitskiy2020image}, and audio \cite{dhariwal2020jukebox}. The present work touches on ways to derive audio embeddings which have supported a variety of applications such as ASR, audio understanding \cite{Chung2018-Speech2Vec}, \cite{verma2020framework}, conditional audio synthesis \cite{skerry2018towards} as well as style, signal transformation \cite{oord2017neural}. We can summarize the contents of the audio signal depending on the task at hand in these small latent representations. Learning a small compressed input representation began the modern deep learning revolution, with the classic work by Hinton \cite{hinton2006reducing}. Once a representation is learned, a classification head similar to \cite{chen2020simple, wang2021multi} is then used to map these vectors to actual labels.
There was a shift to end-to-end neural architectures first in the ASR, by the CLDNN paper proposed by Google in 2015 \cite{sainath2015convolutional}, and then used in tasks like acoustic scene understanding similar to the ImageNet challenge by \cite{aytar2016soundnet}. These architectures quickly surpassed the performance of handcrafted features. \cite{verma2023content} combined the front-end of the work done by raw-CLDNN with the mixture of expert architectures drawing from\cite{jacobs1991adaptive}. This performed better than using a simple convolutional front end with the same Transformer module, showing how elements from traditional signal processing can be combined with classic machine learning ideas. In our work, we provide a direction to improve these architectures by bringing back handcrafted domain-specific features. There have been similar research directions in computer vision, where \cite{jain2022combining} explored diverse sets of feature priors, thus having less overlapping failure modes while dealing with spurious data. However, the goal in our case is different: We do not use them as an ensemble but rather as feature extractors and harness strong inductive domain knowledge to help improve model performance. Before the modern advent of deep learning, several spectro-temporal features were used that could describe characteristics of interest for a particular task. \cite{tzanetakis2002musical} used timbral, energy, rhythm, spectral, frequency-based handcrafted descriptors to identify the contents of the audio signal, in this case, the genre of the music being played. However, end-to-end architectures quickly surpassed them, such as one described in \cite{costa2017evaluation} using convolutional models that could learn features from scratch.
One of the motivations of the current paper is: We typically use data-augmentation\cite{schluter2015exploring} to help with the robustness and scalability of our neural architectures to generalize better to unseen audio/test samples. However, feature-based representations can exist that the model would not encounter in real life. E.g. spectrogram that only contains binary 1/0 mask attributing to the presence/absence of peaks. Or for another case, an MFCC or a neuralogram representation. We cannot reconstruct and get back to the audio signal via these representations. However, they convey a specific meaning/representation for the input signals. Additionally, each one of them is also orthogonal to the other: a binary mask of the location of peaks in a spectrogram only tells us about the location of the frequency content of the audio signal and nothing else. A 13-dim MFCC-based representation gives us only the timbre of the audio signals and nothing more. Thus, we are operating a neural architecture on each of these according to the loss function we used for end-to-end trained architecture and training each separately from scratch. We could have created multiple augmentation schemes and used them in conjunction to train a neural architecture. Another approach by Deepmind explored creating multiple representations of the same audio signal and mapping them to the same latent space \cite{wang2021multi}. However, they do not combine the latent codes but try to make the neural architecture make the latent representation identical or close to each other for unsupervised setups. However, each parameter being learned must consider all the augmentations so that the weights can generalize to unseen scenarios. However, as described earlier, they would only encounter them in the test scenario if we transform the audio in that manner. Hence, we explored the approach described in the current paper. The contributions of our paper are as follows: i) We report how to create feature-based robust neural embeddings for audio signals. These feature-based embeddings are interpretable; for example, for a task, we can see how much pitch-based and timbre-based features would contribute in terms of absolute metrics in accuracy ii) Further, these embeddings are robust; that is, a latent code is learned only by looking at a specific category of features(end-to-end or human-defined), and it uses only those embeddings for a particular task. For example, given a pitch-based representation, it will only use the input provided to learn the best representation for a particular task, unlike passing a raw waveform directly, where it can use any attribute of the signal it deems fit. iii) We showcase how embeddings with prior domain-specific knowledge used with end-to-end architectures can surpass the results obtained using purely learned architecture. This is a very strong finding, as it opens the doors of feature engineering to be used with state of the art architectures. 
\vspace{-0.2cm}
\section{DataSet}
\vspace{-0.2cm}
We work with FSD50K\cite{fonseca2020fsd50k}. This is a classification task: the dataset contains supervised labels, with one or more tags assigned for a specific audio clip. The audio files are variable in length from 1-15s. It contains about 51k audio files, drawn from AudioSet\cite{gemmeke2017audio} ontology. The reader is asked to refer to \cite{fonseca2020fsd50k} for choosing this dataset over AudioSet \cite{gemmeke2017audio}. We primarily chose it for the free availability of the balanced reference dataset, and secondly, being a uniform way of training/testing and reporting results. We only train from scratch neural architecture \textit{only} on this dataset, rather than pre-training on massive audio/vision dataset, unlike \cite{gong2021psla}. We resample all audio files to have a sampling rate of 16kHz. To be consistent with other papers reporting results \cite{fonseca2020fsd50k,verma2021audio,verma2023content}, we do not carry out data augmentation like additive noise, spectral changes, etc. \cite{schluter2015exploring}. All neural architectures are first trained on 1s of audio (similar to the trend started by \cite{fonseca2020fsd50k,gemmeke2017audio} with the architecture predicting one or more labels. The labels of the clip are assigned to each audio chunk of 1s during training, which are learned to be predicted by a neural architecture on representation or from embeddings. The labels averaged across the entire clip to report mean-average precision for clip.
\section{Methodology} We describe the methodology we use to showcase the strength of our work. In most of the literature, data augmentation is used to build robustness into the system, e.g., learning timbral variations, pitch variations, and additive noise, to name a few \cite{schluter2015exploring}. For each input representation, a front-end is defined as going from a feature representation of interest to feed it to the Transformer architecture and adding positional encodings \cite{vaswani2017attention}. The rest of the block remains the same: The Transformer module consists of 6 layers with 64 as the embedding size with a single layer of 256 dimensions acting as an MLP module. We use a dropout rate of 0.3 in the attention and MLP layers, with a 12-number of heads. Global average pooling is carried out at the last layer (6th layer) of the Transformer architecture to get a representation of 64-dim encapsulating the input to get the embedding for the particular input for a particular task. This is consistent with previous work such as \cite{verma2021audio}. Each of the outputs after two Transformer modules is followed by a Max-Pooling block, which reduces the dimension of the number of tokens by a factor of 2. This is successful in computer vision, too, as the final output from the last convolutional layers looks at a much broader receptive field and a hierarchical structure. This is passed onto a linear layer of 2048 neurons followed by a 200-neuron final layer to have the output vector. The loss criterion used to update the weights is Huber loss between the predicted vector of the neural architecture and a 200-dim vector binary vector with 1s present at the location of the category(ies) of the audio present in the input audio representation. All architectures are trained for 300 epochs starting from 2e-4 till 1e-6. For the next subsections, we focus on how to pass on a representation, either end to end or a pre-defined representation-based, onto the Transformer module. Each of these architectures is identical but trained from scratch, except for how to pipe feature-based representation onto the Transformer module. 

\subsection{Frequency Content-Based Representation} Here we only allow pitch/frequency-based information to pass through. Traditionally, pitch detection in a polyphonic setting is a challenging problem. To understand the frequency content, we do not want any other information like the energy and timbre of the signal in our representation. For each 1s audio chunk, we first compute a log-magnitude constant-Q representation \cite{brown1991calculation} with the hop length of 25ms, for 80 bins, with 12 bins for every octave doubling starting from 40Hz, and a sparsity factor chosen to be 0.01 using Librosa library \cite{mcfee2015librosa}. We only retain spectral peaks, with peaks picked in individual slices, by looking at +/-2 spectral bins of either side. Further, we only retain the peaks of absolute strength greater than equal to the median of the log-magnitude of the contents in the 1s of the constant-Q representation. This will retain the spectral/harmonic structure of the contents of the audio signal yet will only have binary values that correspond to the presence/absence of peaks. The front-end encoder, in this case, takes an 80-dim vector corresponding to the single slice of our representation, learns a 64-dim embedding to conform it to the embedding dimension of the Transformer.
\subsection{Timbre Based Representation} To represent timbre-based information, we compute a 13-dim MFCC \cite{logan2000mel} representation and throw away the first coefficient to get a 12-dimension vector every 25ms to get a representation of MFCC coefficients of dimension 12x40 time steps. This is piped through a front end similar to the frequency-based content representation, i.e., projected to a dimension of 64-dim via a linear layer, and sinusoidal positional embeddings added before 1st Transformer layer.  
\subsection{End-To-End Architecture} The recipe for an end-to-end Transformer follows the classical work of \cite{sainath2015learning}, and more recently, that of \cite{verma2023content}. We divide an input waveform into patches of 25ms, thus having 40 chunks. Each 25ms comprising 400 samples is passed through a series of convolutional filters of length 200, with the number of filters being 128, with zero-padded such that the output of each convolutional filter is the same length as the input. We take the maximum across the output of the convolutional filter for each convolutional filter to get a single vector of length equal to number of conv filters for each 25ms. This vector is now projected to a dimension of 64 via a linear layer, and sinusoidal positional embedding is added before being passed onto the Transformer module.
\subsection{Neuralogram: Stacked Embeddings} In this representation, we project each of the 100ms waveform chunks through a convolutional architecture, MobileNet \cite{howard2017mobilenets} trained on FSD-50K with same way as our baseline architecture. We choose this as it is much more efficient on performance per number of parameters than other neural architectures. This representation of stacking the output of the last convolutional layer (Neuralogram \cite{verma2019neuralogram}), when we pass on an input of 100ms, gives a 1024-embedding vector, gives us a representation of 1024x10 shape for 1s content of the audio signal. This vector is now projected to a dimension of 64 via a linear layer, and sinusoidal positional embedding is added before being passed onto the Transformer module. This differs from an end-to-end architecture, as the architectures are first different. Secondly, we use the convolutional module only as a projector of smaller waveform chunks, and a Transformer architecture still learns the actual dependencies across time. Individually, understanding the contents of the signal, with just 100ms, is a difficult task, and generally, for humans, too, more context is needed. However, the embeddings are projected onto a space that can be utilized/fed to the Transformers to understand the context. We train each of them on the embeddings separately with the given labels and combine them to get an interpretable and robust stacked representation, achieving significant gains when used together. 
\vspace{-0.2cm}
\begin{table}[ht]
  \caption{Feature Representation(Top-5 Accuracy) on 1s Test set of FSD-50K}
	\centering
	\begin{tabular}{|c|c|c|}
		\hline
		Feature Set Used  &Accuracy\\\hline
		Pitch Representation  &  27.2 \%\\
        MFCCs - Timbre Representation & 34.2 \% \\
       \textbf{Neuralogram-Convolutional Embedding} & 43.7\% \\
       Audio Transformer -- End-to-End &  41.9\%\\
       \hline
       \hline
       Linear Model on Diverse Embeddings & \textbf{44.9} \%\\
		 \hline
	\end{tabular}
	\label{tab:example}
\end{table}
\vspace{-0.2cm}
\subsection{Combining Embeddings} For the baseline architecture, we report the top-5 accuracy, from the test set for an end-to-end architecture. This is the exact model as described in the section above. As per our introduction, we use the embedding of dimension 64 dimensions to get the representation or embedding for each one of the feature sets, namely i) pitch/frequency content, ii) timbre, iii) end-to-end architecture, and iv) pretrained stacked embedding. Since we train each of these four architectures separately, on the input as described before, all of which correspond to 1s of audio, we treat the 64-dim output before the linear classification head (after taking the global average pooling operation) as the representation/embedding for the contents of that audio signal. We now train the same linear classifier onto the subset of the embeddings to see how well we do. We stack a robust, diverse, interpretable feature set that makes sense together with an end-to-end learned architecture.
\vspace{-0.2cm}
\section{Results and Discussion} We, in the first experiment, report how well we do when we use each of the input feature representations or end-to-end architectures individually. We trained each model architecture separately and reported the top-5 accuracy on FSD-50K in Table 1. The best results are obtained by learning an embedding by projecting waveform onto MobileNet embeddings and training a 6-layer Transformer module. This method surpassed an end-to-end learned architecture with the same number of parameters. One hypothesis is that most heavy lifting has already been carried out by conv net to project waveform patches to separable neural embeddings. Hence, most of the parameters of neural transformers are dedicated to learning interdependencies amongst the latent codes, as opposed to the end-to-end model, which has to learn the separable latent codes and the connections from scratch. The important point is that a pitch-based representation of retaining just the binary locations of spectral peaks does not achieve competitive results. However, the embedding learned is robust, as the model tries its best to achieve the best results in a sub-optimal representation without taking help from other features that an end-to-end or a waveform-based architecture might take. Similar arguments can be made for other input representations. This work shows that learning diverse latent embeddings on a (sub-optimal) representation and then re-training just the linear head achieves a state-of-the-art accuracy on datasets without using extra training data. We also see that we get a significant bump in performance as compared to a audio Transformer trained on raw waveforms. Further, by utilizing our proposed algorithm, we outperform CLAP a popular algorithm, and PLSA trained on extra training data  reinforcing the strength of our work. We do not achieve state of the art performance, as we do not use extra training data. Further the number of parameters of our architecture ~4 million are miniscule in terms of 40 billion parameter architectures such as One-Peace that are trained on massive amounts of open-source audio sources. This further shows, how with interpretable features grounded in signal processing ideas deserve further explorations.
\begin{table}[t]
  \caption{ Comparison of Mean-Average Precision with other architectures. }
	\centering
	\begin{tabular}{|c|c|c|}
		\hline
		Neural Model Architecture & extra data & MAP\\\hline
		DenseNet \cite{fonseca2020fsd50k} & No & 42.5\\
		Audio-Transformer \cite{verma2021audio} & No & 53.7\\
		Knowledge Distillation \cite{choi2022temporal} & No & 54.8\\
        Wav2CLIP \cite{wu2022wav2clip} & No &  43.1 \\
        Bank of Filterbanks \cite{verma2023content} & No & 55.2 \\
        PLSA \cite{gong2021psla} & Yes & 56.7 \\
        PaSST-N-S \cite{koutini2021efficient} & Yes & 64.2 \\
        One-Peace \cite{wang2023one} & Yes &  69.7 \\
        CLAP \cite{elizalde2023clap} & Yes &  58.6 \\\hline
        %Our baseline  waveform model & No & 56.5\\
        \textbf{Our LM on Diverse Embeddings} & No & \textbf{59.6}\\\hline
		
	\end{tabular}
	\label{tab:example}
\end{table}
\section{Conclusion and Future Work}
This paper shows how prior domain-specific feature embeddings can be extracted and used in conjunction with end-to-end learned embeddings. This is particularly important, as by learning feature-specific embeddings, we learn a robust feature set that focuses on the best representation in that domain-specific representation rather than taking help from other signals. In this work, we see that we diversify our feature set by first learning a diverse feature set based on pitch, timbre, end-to-end architecture, and convolutional embedding. These category-specific features, combined with end-to-end architecture-derived embedding, not only add to the interpretability and robustness of the learned representation but also help us increase the performance of a baseline end-to-end learned architecture by quite a significant amount. We hope this work will pave the way for bringing in domain expertise and optimized end-to-end architectures.

% -------------------------------------------------------------------------
% Either list references using the bibliography style file IEEEtran.bst
% The \IEEEtriggeratref{XX} command can be used to move to the next column before the XX-th reference
% to balance the two columns of the reference section
% \IEEEtriggeratref{XX}
\bibliographystyle{IEEEtran}
\bibliography{refs25}
% or list them by yourself:
% \begin{thebibliography}{1}

% \bibitem{waspaaweb}
% {WASPAA Website}, \url{http://www.waspaa.com}.

% \bibitem{IEEEXploreReqs}
% {IEEE {X}plore {R}equirements}, \url{https://conferences.ieeeauthorcenter.ieee.org/write-your-paper/meet-ieee-xplore-requirements/}.

% \bibitem{eWilliams1999}
% E.~Williams, \emph{Fourier Acoustics: Sound Radiation and Nearfield Acoustic Holography}.\hskip 1em plus 0.5em minus 0.4em\relax London, UK: Academic Press, 1999.

% \bibitem{cJones2003}
% C.~Jones, A.~Smith, and E.~Roberts, ``A sample paper in conference proceedings,'' in \emph{Proc. ICASSP}, vol.~II, Apr. 2003, pp. 803--806.

% \bibitem{aSmith2000}
% A.~Smith, C.~Jones, and E.~Roberts, ``A sample paper in journals,'' \emph{IEEE Trans. Signal Process.}, vol.~62, pp. 291--294, Jan. 2000.

% \end{thebibliography}

\end{document}